\documentclass[12pt,preprint]{aastex}

\def\arcsecpoint{$''\!.$}
\def\deg{$^{\rm o}$}
\def\ltsim{\raisebox{-.5ex}{$\;\stackrel{<}{\sim}\;$}}
\def\gtsim{\raisebox{-.5ex}{$\;\stackrel{>}{\sim}\;$}}

\shortauthors{Crenshaw et al.}
\shorttitle{UV Absorption in NGC~4395}

\begin{document}

\title{High-Resolution Ultraviolet Spectra of the Dwarf Seyfert 1 Galaxy NGC 
4395: Evidence for Intrinsic Absorption\altaffilmark{1}}

\author{D.M. Crenshaw\altaffilmark{2},
S.B. Kraemer\altaffilmark{3},
J.R. Gabel\altaffilmark{4},
H.R. Schmitt\altaffilmark{5},
A.V. Filippenko\altaffilmark{6},
L.C. Ho\altaffilmark{7},
J.C. Shields\altaffilmark{8},
and T.J. Turner\altaffilmark{9}
}

\altaffiltext{1}{Based on observations made with the NASA/ESA {\it Hubble 
Space Telescope}, obtained at the Space Telescope Science Institute, which is 
operated by the Association of Universities for Research in Astronomy, Inc., 
under NASA contract NAS 5-26555; these observations are associated with 
proposal GO-9362.
Also based on observations made with the NASA-CNES-CSA {\it Far Ultraviolet 
Spectroscopic Explorer}. {\it FUSE} is operated for NASA by Johns Hopkins 
University under NASA contract NAS5-32985.}

\altaffiltext{2}{Department of Physics and Astronomy, Georgia State 
University, Astronomy Offices, One Park Place South SE, Suite 700,
Atlanta, GA 30303; crenshaw@chara.gsu.edu.}

\altaffiltext{3}{Catholic University of America and Laboratory for Astronomy
and Solar Physics, NASA's Goddard Space Flight Center, Code 681,
Greenbelt, MD  20771; stiskraemer@yancey.gsfc.nasa.gov.}

\altaffiltext{4}{Center for Astrophysics and Space Astronomy, University of 
Colorado, 389 UCB, Boulder, CO 80309-0389; Jack.Gabel@colorado.edu.} 

\altaffiltext{5}{National Radio Astronomy Observatory, 520 Edgemont Road, 
Charlottesville, VA 22903; hschmitt@nrao.edu.}

\altaffiltext{6}{Department of Astronomy, University of California, Berkeley, 
CA 94720-3411; alex@astro.berkeley.edu.}

\altaffiltext{7}{The Observatories of the Carnegie Institution of Washington, 
813 Santa Barbara Street, Pasadena, CA 91101-1292; lho@ociw.edu.}

\altaffiltext{8}{Department of Physics and Astronomy, Ohio University, Athens, 
OH 45701-2979; shields@helios.phy.ohiou.edu.}

\altaffiltext{9}{Joint Center for Astrophysics, University of Maryland, 
Baltimore County, 1000 Hilltop Circle, Baltimore, MD 21250;
turner@lucretia.gsfc.nasa.gov.}

\begin{abstract}

We present ultraviolet spectra of the dwarf Seyfert 1 nucleus of NGC~4395, 
obtained with 
the {\it Far Ultraviolet Spectroscopic Explorer (FUSE)} and the {\it Hubble 
Space Telescope's} Space Telescope 
Imaging Spectrograph at velocity resolutions of 7 to 15 km s$^{-1}$. We 
confirm our earlier claim of C~IV absorption in low-resolution UV spectra and 
detect a number of other absorption lines with lower ionization 
potentials. In addition to the Galactic lines, we identify two kinematic 
components of absorption that are likely to be intrinsic to NGC 4395.
We consider possible origins of the absorption, including 
the interstellar medium (ISM) of NGC 4395, the narrow-line region (NLR), 
outflowing UV absorbers, and X-ray ``warm absorbers.''
Component 1, at a radial velocity of $-$770 km s$^{-1}$ with respect to the 
nucleus, is only identified in the C~IV $\lambda$1548.2 line. It most likely 
represents an outflowing UV absorber, similar to those seen in a majority of 
Seyfert 1 galaxies, although additional observations are needed to confirm the 
reality of this feature. Component 2, at $-$114 km s$^{-1}$, most likely arises 
in the ISM of NGC~4395; its ionic column densities cannot be matched by 
photoionization models with a power-law continuum.
Our models of the highly ionized X-ray absorbers claimed for this active 
galactic nucleus 
indicate that they would have undetectable C~IV absorption, but large O~VI and 
H~I columns should be present. We attribute our lack of detection of the O~VI 
and Ly$\beta$ absorption from the X-ray absorbers to a combination of noise and 
dilution of the nuclear spectrum by hot stars in the large {\it FUSE} aperture.

\end{abstract}

~~~~~

\section{Introduction}

A majority ($\sim$60\%) of Seyfert 1 galaxies show outflow of ionized gas
from their nuclei, as revealed by intrinsic ultraviolet (UV) and X-ray 
absorption lines that 
are blueshifted with respect to the systemic velocities of the galaxies.
The significance of the outflows is illustrated by the fact that the inferred 
mass-loss rates are comparable to the mass-accretion rates (Crenshaw, Kraemer, 
\& George 2003a). 
Dynamical models of the outflowing absorbers make use of thermal winds, 
radiation pressure, and/or hydromagnetic flows from the central accretion disk, 
presumably responsible for most of the continuum radiation, or from the nearby 
broad-line region or torus.
To constrain the dynamical models, and 
in particular to test the hypothesis that radiation pressure is the driving 
force, it is important to determine their rate of occurrence and variation in 
properties as functions of the luminosity of the central continuum source.

At luminosities higher than those of Seyferts, a significant fraction of 
quasars (25\% to 55\%) show intrinsic absorption in the UV (Ganguly et al. 
2001; Laor \& Brandt 2002; Vestergaard 2003) and X-rays (George et al. 2000). 
However, the case for intrinsic absorption in low-luminosity active galactic 
nuclei (AGN) has not been 
firmly established. Shields et al. (2002) find that LINERs often show 
absorption lines, but these can usually be attributed to the interstellar 
medium (ISM) in their host galaxies, and there is little evidence in most cases 
for absorbers outflowing from their nuclei.

To date, NGC~4395 provides the best case for a low-luminosity AGN with
intrinsic absorption that could possibly be attributed to the nucleus. 
NGC~4395 is an Sd III-IV dwarf galaxy that harbors one of the nearest ($d 
\approx 4.6$ Mpc, Karachentsev et al. 2003) and least luminous ($L_{bol} 
\approx 5 \times 10^{40}$ ergs s$^{-1}$, Moran et al. 2004) AGN 
known. The active nucleus can be thought of as a dwarf Seyfert 1, since it 
possesses a bright central continuum source that is rapidly variable in the 
X-rays (Lira et al. 1999; Moran et al. 1999, 2004; Shih et al. 2003, hereafter 
SH2003) as well as 
broad and narrow emission lines that are photoionized by the central source 
(Kraemer et al. 1999). A variety of evidence indicates that the nucleus harbors 
a relatively low mass (10$^{4}$  -- 10$^{5}$ M$_{\odot}$) black hole that is 
responsible for powering the nuclear activity (Filippenko \& Ho 2003). The 
broad permitted lines,
strong high-ionization lines (Kraemer et al. 1999), and 
relatively high accretion rate ($L_{bol}/L_{Edd} \approx 2 - 20
\times 10^{-3}$) compared to that of LINERs ($L_{bol}/L_{Edd} < 10^{-3}$, see 
Ho 1999) strongly support the Seyfert 1 interpretation for NGC~4395.

Several previous studies of NGC~4395 have made a case for intrinsic UV or X-ray 
absorption from ionized gas in the line of sight to the nucleus. Faint Object 
Spectrograph (FOS) observations obtained at low spectral resolving power 
($\lambda$/$\Delta\lambda$ $\approx$ 1000) with the {\it Hubble Space 
Telescope} ({\it HST}) show absorption features in the blue wing of the C~IV 
emission line, at the observed wavelengths of 1544 and 1550 \AA\ (Filippenko, 
Ho, \& Sargent 1993; Kraemer 
et al. 1999). Early X-ray observations with {\it ROSAT} (Lira et al. 1999; 
Moran et al. 1999) and {\it ASCA} (Iwasawa et al. 2000) showed evidence for 
absorption in soft X-rays, indicative of one or more X-ray ``warm 
absorbers.'' Subsequent {\it Chandra X-ray Observatory} ({\it CXO}) (Moran et 
al. 2004) and long {\it ASCA} (SH2003) observations showed a
downturn in the X-ray spectrum at $E $\ltsim$ 3$ keV, apparently due to 
absorption from highly ionized gas.

In an effort to confirm the presence of intrinsic UV absorption in NGC 4395, 
and determine its properties and connection with the claimed X-ray absorption, 
we have obtained new observations with the {\it Far Ultraviolet 
Spectroscopic Explorer (FUSE)} and the Space Telescope Imaging Spectrograph 
(STIS) on {\it HST}. We were also motivated by our previous study of the 
emission-line regions in NGC 4395 (Kraemer et al. 1999), which revealed that 
both the broad-line region (BLR) and narrow-line region (NLR) should have large 
($\gtsim$50\%) covering factors of the central source, and thus might be 
detectable in absorption.

\section{Observations}

We observed the nucleus of NGC 4395 contemporaneously with {\it FUSE} and STIS 
at high spectral resolutions, to identify possible absorption systems in the 
UV. We obtained the {\it FUSE} observations on 2003 February 25 UT with a total 
exposure time of 36,000 s. We used the standard 
30$''$ $\times$ 30$''$ low-resolution aperture (LWRS), which samples a much 
larger circumnuclear region than our STIS observation (see below). We 
processed the {\it FUSE} data with version 2.2.3 of the standard calibration 
pipeline, CALFUSE, which extracts spectra for each of the eight combinations 
of channels (SiC1, SiC2, LiF1, LiF2) and detector segments (A and B). 
Curvature in the LiF1B spectrum due to the anomalous feature 
known as ``the worm'' (Sanhow et al. 2002) was corrected by comparison
to the LiF2A spectrum.  The channel/segments with overlapping wavelength 
coverage with the LiF1a spectrum were scaled to match its flux.
All spectra were then co-added by weighting each channel/segment by 
its effective area, using the effective area versus wavelength 
functions given in Blair et al. (2000). The final calibrated spectrum covers 
the range 905 -- 1180 \AA\ at a velocity resolution of $\sim$15 km s$^{-1}$ 
(full-width at half-maximum [FWHM] of the line-spread function).

We obtained STIS echelle spectra of the nucleus of NGC~4395 through the 
0\arcsecpoint2 $\times$ 0\arcsecpoint2 aperture on 2003 March 9 UT. We used the 
E140M grating over three consecutive {\it HST} orbits to yield a total exposure 
time of 8022 s; the spectra cover the range 1150 -- 1730 \AA\ at a velocity 
resolution of $\sim$7 km s$^{-1}$ (FWHM). We also obtained a 2182 sec exposure 
with the E230M grating during one orbit to cover 
the range 2271 -- 3119 \AA\ at a velocity resolution of $\sim$10 km s$^{-1}$ 
(FWHM). We reduced the STIS spectra using the IDL software developed at NASA's 
Goddard Space Flight Center for the STIS Instrument Definition Team.
The data reduction included a procedure to remove the background light from 
each order using a scattered light model. The individual orders in each 
echelle spectrum were spliced together in the regions of overlap.

We measured the continuum fluxes of the three individual STIS E140M spectra 
and found no significant ($\geq$ 10\%) variations on short time scales (i.e., 
from one $\sim$90 min orbit to the next). The average and standard deviation 
of the continuum fluxes at 1345~\AA\ (in a bin of 30 \AA) are 1.55 
($\pm$ 0.11) $\times$ 10$^{-15}$ ergs s$^{-1}$ cm$^{-2}$ \AA$^{-1}$. This 
happens to 
be close to the 1345~\AA\ continuum flux from the FOS spectrum obtained on 
1992 July 15, 19 (Filippenko, Ho, \& Sargent 1993): 1.41 $\times$ 10$^{-15}$ 
ergs 
s$^{-1}$ cm$^{-2}$ \AA$^{-1}$. Given the lack of obvious variability on 
90-minute time scales, we averaged the E140M spectra together to improve the 
signal-to-noise ratio.

\section{Hot Stars in the FUSE Aperture}

Comparison of the STIS and {\it FUSE} spectra in the region of overlap (1150 -- 
1180 \AA) reveals that the {\it FUSE} continuum flux is 3.0 times higher than 
that from STIS. Although we cannot rule out variability in the UV continuum 
flux over the 12-day interval between the 
observations, we demonstrate that the principal source of the flux discrepancy 
is the presence of hot stars in the large {\it FUSE} aperture. This is not 
surprising, since there is intense star formation occurring throughout 
NGC 4395 (Cedr\'es \& Cepa 2002).

Given that there are no available far-UV images of NGC 4395 (which will change 
with the release of the GALEX all-sky survey), we test the hot-star
hypothesis in three ways: with a near-UV image, with the H$\alpha$
images of Cedr\'es \& Cepa (2002), and by comparing the FUSE spectrum
with starburst models from Leitherer et al. (1999). We have examined a
near-UV image (centered at 3300\AA) obtained with the Advance Camera
for Surveys (ACS) on {\it HST} on 2002 October 27 UT. The integrated flux in 
the 25$''$ $\times$ 28$''$ ACS image, which is close to the projected area of 
the {\it FUSE} aperture, is 3.1 times that measured in a region corresponding 
to the STIS aperture (H.R. Schmitt et al. 2003, in preparation), in close 
agreement with the far-UV ratio measured from the spectra.

Cedr\'es \& Cepa (2002) detected three H$\alpha$ sources inside the region 
covered by the {\it FUSE} aperture: the nucleus, a source $\sim$7\arcsec\ to 
the E, and another source $\sim$5\arcsec\ to the S of the nucleus (regions 99, 
98 and 97, respectively, in their paper). Assuming that all the nuclear 
H$\alpha$ emission is ionized by the AGN, we get that $F$(H$\alpha$) $=$ 4.45 
$\times$ 10$^{-14}$ ergs s$^{-1}$ cm$^{-2}$
for the other two sources. Based on this value, we calculate
from equation (2) in Kennicutt (1998) that this corresponds to a star
formation rate of 8.4 $\times$ 10$^{-4}$~M$_{\odot}$~yr$^{-1}$. This star
formation rate corresponds to a far-ultraviolet flux of $F$$_{\lambda}$(1150 
\AA) $=$ 5.7 $\times$ 10$^{-15}$ ergs s$^{-1}$ cm$^{-2}$ \AA$^{-1}$ using
equation (1) in Kennicutt (1998).
These calculations are based on a continuous star formation rate over
a period of $\sim$100~Myr, which can give different results
from the ones obtained using an instantaneous burst. According to
Cedr\'es \& Cepa (2002), the equivalent widths of H$\alpha$ emission
are log (EW/\AA) $=$ 2.46 and 2.61 for regions 97 and 98, respectively.
Using the Starburst99 models for an instantaneous burst with metallicity
0.4 solar and Salpeter IMF (Leitherer et al. 1999), we find that these
equivalent widths correspond to bursts with ages of $\sim$5~Myr. Scaling
the 5~Myr model to match the observed H$\alpha$ flux, we get a 
continuum flux of $F$$_{\lambda}$(1150 \AA) $=$ 8.5 $\times$ 
10$^{-15}$ ergs s$^{-1}$ cm$^{-2}$ \AA$^{-1}$. Both of these estimates are 
close to the excess far-UV flux in the {\it FUSE} aperture of
F$_{\lambda}$(1150~\AA) $=$ 5.9 ($\pm$1.1) $\times$ 10$^{-15}$ ergs s$^{-1}$ 
cm$^{-2}$ \AA$^{-1}$.

The third test for the hot stars hypothesis was done by comparing the
{\it FUSE} spectrum with starburst models created using Starburst99 (Leitherer
et al. 1999), which incorporate far-UV stellar spectra from Robert et al.
(2003). The models assume an instantaneous burst with
a Salpeter IMF, upper and lower mass limits of 100~M$_{\odot}$ and
1~M$_{\odot}$, respectively, and a metallicity of 0.4Z$_{\odot}$, which is
consistent with the values measured by Cedr\'es \& Cepa (2002). We 
restrict our comparison of the FUSE spectrum to models with ages between 3~Myr
and 7~Myr, which is consistent with the age obtained for regions 97 and 98
based on the equivalent widths of H$\alpha$.
In order to compare the FUSE spectrum with the starburst models, we first
corrected the spectrum of NGC 4395 for Galactic reddening,
using the value E(B-V) $=$ 0.074 from Schlegel, Finkbeiner \& Davis (1998).
Internal reddening is not a problem in the nucleus of this galaxy
(Ho, Filippenko \& Sargent 1997). We then corrected the spectrum for redshift, 
and resampled it to 0.13 \AA\ in wavelength, which is the spectral resolution
of the models.

The resulting spectrum is presented as a thin line in Figure 1.
Since approximately one third of the flux observed by FUSE at 1150 \AA\
originates at the nucleus, we used the $HST$ spectrum to determine the slope
of the AGN continuum and extrapolate it to lower wavelengths. We did this by
fitting a power law (F$_{\lambda}\propto\lambda^{\alpha}$) to the continuum 
regions in the wavelength range 1268 -- 1730 \AA\ (thereby avoiding possible 
contamination from stars at longer wavelengths). We found
that the AGN continuum can be represented by a power law with slope
$\alpha=-0.59\pm0.58$. The extrapolation of this power law to shorter
wavelengths is shown as a thick dotted line in Figure 1. We also show a 
spectrum from a 5~Myr starburst model (thick solid line), scaled so that when
added to the power law, its flux is the same as the one observed in
NGC 4395 over the wavelength range 1102-1107\AA.
Due to the low S/N ratio of the {\it FUSE} spectrum of NGC\,4395, we did
not attempt a detailed modeling of the stellar absorption lines; comparison 
with starburst spectra with ages between 3~Myr and 7~Myr did not show a 
significant difference, so we adopted the model which
corresponded to the age determined from the H$\alpha$ equivalent widths.
Most of the absorption lines detected in the spectrum are due to interstellar 
gas in our Galaxy or NGC 4395. However, several stellar absorption lines, 
notably from C~III, S~IV, P~V and Si~IV, are clearly present (we could not 
detect O~VI 
absorption because it is filled in with emission from the AGN). The strengths 
of these lines are similar in NGC 4395 and in the model spectrum. A subtraction 
of the starburst plus power law model from the spectrum of NGC~4395 (bottom 
panel of Figure 1) successfully removes the strongest stellar and large-scale 
continuum features, leaving residuals primarily due to interstellar absorption 
and nuclear emission lines. Based on all the tests presented in this section, 
we conclude that a starburst is the correct explanation for
the excess emission observed by FUSE. This turns out to have 
important consequences for the detection of intrinsic absorption in the {\it 
FUSE} aperture (\S5.3).

\section{Detection and Measurement of Intrinsic UV Absorption}

Due to the relative faintness of NGC 4395, its STIS spectra are 
noisier than the STIS echelle spectra of other Seyfert galaxies that we have 
studied in the past (Crenshaw et al. 2003a). The signal-to-noise ratio
per resolution element is S/N $=$ 2.3 in the regions of broad C~IV emission 
surrounding the intrinsic absorption, whereas S/N $\approx$ 8 in these regions 
for similar exposure times in NGC~5548 (Crenshaw et al. 2003b). For the {\it 
FUSE} spectra, we estimate S/N $\approx$ 3 (per resolution element) in the 
O~VI region. Since intrinsic absorption lines are expected to encompass many 
resolution elements (Crenshaw et al. 1999), we smoothed the spectra by a 
7-point boxcar function to search for absorption lines.

Figure 2 shows portions of the STIS echelle spectra in regions where we have 
detected absorption lines that are likely to be intrinsic to NGC~4395. From the 
narrow emission lines of the C~IV doublet seen in the top panel, we have 
determined an emission-line redshift of $z =$ 0.0012, close to the value of 
0.0011 determined from H~I 21-cm emission by Haynes et al. (1998). We note that
these emission lines are unusually narrow compared to those of normal Seyfert 
galaxies; for C~IV the FWHM is 65 km s$^{-1}$, similar to the value of 
$\sim$50 km~s$^{-1}$ determined for various narrow emission lines from a 
high-resolution optical 
spectrum of NGC~4395 (Filippenko \& Ho 2003). Unfortunately, the N~V 
$\lambda\lambda$1238.8, 1242.8 region is very noisy, and no absorption (not 
even Galactic) was detected in this region. Strong, broad Galactic Ly$\alpha$ 
absorption prevents detection of any intrinsic Ly$\alpha$ absorption.

Most of the absorption lines that we have detected in the STIS spectra can be 
attributed to the Galaxy, at a radial velocity of $-$29 km s$^{-1}$ in the 
observed frame ($-$389 km s$^{-1}$ in Figure 2). However, we have 
identified two possible kinematic components 
of absorption that could be intrinsic to NGC~4395. Component 1 at $-$770 km 
s$^{-1}$ (with respect to $z =$ 0.0012) is only evident in the C~IV 
$\lambda$1548.2 line. The associated C~IV $\lambda$1550.8 line is 
blended with Galactic absorption, so we regard this identification as 
tentative. There is little or no low-ionization gas associated with this 
component; in particular, there is no obvious Mg~II absorption. Component 2 at 
$-$114 km s$^{-1}$ is seen in the lines of C~IV $\lambda\lambda$1548.2, 1550.8,
Si~IV $\lambda\lambda$1393.8, 1402.8, Mg~II $\lambda\lambda$2796.3, 2803.5, 
and O~I $\lambda$1302.2 (also C~II $\lambda$1334.5, which is not shown).
Although the lines from Component 2 do not align perfectly, it is not possible 
to determine if this is due to ion-dependent velocity structure or just noise.
Component 2 appears to be heavily saturated in all lines except for C~IV. In 
the top panel of Figure 2, we have overplotted the 1992 FOS spectrum (corrected 
to match the STIS spectrum by adding 2 \AA\ to the wavelength scale). The two 
broad dips in the FOS spectrum correspond to our claims of absorption features 
at 1544 and 1550 \AA\ (Filippenko, Ho, \& Sargent 1993; Kraemer et al. 1999). 
These can now be attributed to blends of the intrinsic and Galactic absorption 
lines that we have identified in the STIS spectrum, plus some contribution of 
the gap between the narrow C~IV emission doublet to the dip at 1550 \AA\ ($+$ 
300 km s$^{-1}$ in the plot).

Figure 3 shows portions of the {\it FUSE} spectra. Due to the low S/N at 
short wavelengths, we were unable to detect even Galactic absorption in the 
expected lines of C~III $\lambda$977.0 and N~III $\lambda$989.8. However, 
Ly$\beta$ and C~II $\lambda$1036.3 are clearly present in Component 2. The 
Ly$\beta$ absorption at Component 2 is not just the wing of the Galactic 
component, since 
symmetric absorption is not present on the other (more negative) side of 
geocoronal Ly$\beta$ emission. As the top panel of Figure 3 shows, we have not 
detected O~VI absorption in any component, despite the reasonably decent S/N 
(the absorption near the expected locations of Components 1 and G in O~VI 
$\lambda$1037.6 is due to C~II, as shown in the bottom panel).

To further characterize the absorption components, we have measured their 
radial velocity centroids and equivalent widths (EWs) assuming full coverage 
of the continuum and line emission. We determined uncertainties in the 
absorption measurements from photon statistics and different reasonable 
placements of the continuum and emission-line profiles. Since these data are 
too noisy to determine covering factors and ionic column densities from the 
optical depths of the lines, we have estimated column densities for Components 
1 and 2 from the EWs of the lines. As mentioned previously, 
nearly all of the detected lines appear to be saturated, so we used the EWs to 
calculate lower limits to their columns. Since C~IV does not appear to be 
saturated in Component 2, we give an actual value and estimated error for its 
column. Note that if the covering factor for C~IV is significantly less than 
one, which seems unlikely given the depths of the other lines in this 
component, the quoted value for the C~IV column density is a lower limit.

Our measurements are summarized in Table 1. In addition to the 
velocity centroids and EWs for Component 2, we list the corresponding Galactic 
values for comparison. For the Galactic component, the average velocity 
centroid is $-$29 ($\pm$20) km s$^{-1}$ in the observed frame. For Component 
2, the average velocity centroid is $-$114 ($\pm$21) km s$^{-1}$ with respect 
$z =$ 0.0012, and the FWHM of the unsaturated C~IV absorption is 60 ($\pm$22) 
km s$^{-1}$. If Component 1 is real and outflowing with respect to z $=$ 
0.0012, we estimate a radial velocity and FWHM of $-$770 ($\pm$21) and 130 
($\pm$30) km s$^{-1}$, respectively. An EW of 0.84 ($\pm$0.21) \AA\ gives a 
lower limit of 2.0 ($\pm$ 0.5) $\times$ 10$^{14}$ cm$^{-2}$ for the C IV column 
density in Component 1.

\section{Origin of the Absorption}

In general, the absorption components that we have detected could arise in 1) 
our Galaxy (disk or halo), 2) the ISM of NGC~4395, 3) the NLR of NGC~4395, 4) 
outflowing UV absorbers similar to those found in most Seyfert 1 galaxies, or 
5) X-ray absorbers, characterized by higher ionization parameters than those of 
typical UV absorbers (Crenshaw et al. 2003a). We detect a number of 
absorption lines from the Galaxy 
(Component ``G'') at an observed radial velocity of $-$29 km s$^{-1}$; their 
EWs are similar to those found in other studies of Galactic absorption (Savage 
\& Sembach 1994). Since the redshift of NGC 4395 is low, and absorption lines 
from the Galactic halo can appear at radial velocities of several hundreds of 
km s$^{-1}$ in the observed frame, we need to consider the possibility of a 
Galactic origin for the other absorption components. We rule out a Galactic 
origin for Component 2, at a radial velocity of $+$246 km s$^{-1}$ in the 
observed frame, since at the Galactic coordinates for NGC~4395 ($l = 
162.2$\deg, $b = 81.5$\deg), all of the high-velocity HI (Wakker 1991) and O~VI 
(Sembach et al. 2003) clouds detected in this area 
of the sky have negative radial velocities (Wakker 1991). Component 1 is at a 
radial velocity $-$410 km s$^{-1}$ in the observed frame; although it could 
conceivably originate in the Galactic halo, its velocity is at least 100 km 
s$^{-1}$ more negative than other H~I and O~VI halo clouds in this area of the 
sky (Wakker 1991; Sembach et al. 2003). Also, we would expect to see detectable 
Si~IV for this component if it 
were halo gas (Savage \& Sembach 1994), and none is apparent (see Figure 2). 
Thus, it is likely that these two components do not arise from the Galaxy.

Given an intrinsic origin for Components 1 and 2, the distinction between the 
above sources may seem somewhat ambiguous. For example, some outflowing UV 
absorbers are likely a component of the NLR (Kraemer et al. 2001), while others 
have been linked to X-ray absorbers, such that the UV absorption lines are due 
to trace elements in the highly ionized, high-column gas (Mathur et al. 1994; 
Kraemer et al. 2002). For this discussion, the intrinsic absorption 
components are placed into the general category of outflowing UV absorber if 
they are not consistent with an origin in the other categories, which have been 
well characterized in NGC 4395 by other studies, as discussed in the following 
subsections.

\subsection{Component 1}

First, we need to consider whether or not Component 1 is ``real,'' 
given the detection of one C~IV absorption line in noisy data. On the positive 
side, there is an inflection in the FOS spectrum at this position that appears 
to be distinct from the inflection caused by the Galactic C~IV $\lambda$1548.2 
line (Figure 2). On the negative side, no other lines have been detected, and 
in particular, Ly$\beta$ appears to be absent at the radial velocity of 
Component 1. We have been unable to devise a photoionization model that would 
result in an undetectable Ly$\beta$ line, since even very highly ionized gas 
will have a substantial H~I column. Assuming the C~IV absorption is real, the 
associated Ly$\beta$ absorption must be diluted by unabsorbed UV continuum 
emission in the {\it FUSE} aperture, which we characterized in \S3.

The high radial velocity of Component 1 ($-$770 km s$^{-1}$) with respect to 
NGC~4395 rules out an origin in the ISM of this galaxy. As can be seen in 
Figures 2 and 3, the wings of the narrow components of the 
emission lines do not extend past about $\pm$100 km s$^{-1}$ from line center,
so an origin in the NLR is ruled out as well. As discussed in \S5.3, the X-ray 
absorbers characterized by SH2003 are too highly ionized to produce significant 
columns of C~IV or lower-ionization species. Thus, Component 1 likely 
originates in an outflowing UV absorber, which is defined by the presence of 
C~IV absorption that is blueshifted with respect to the systemic velocity of 
the host galaxy (Crenshaw et al. 1999). The absence of Si~IV and lower 
ionization 
lines is not a problem for this interpretation, since most UV absorbers are too 
highly ionized to have substantial columns of these ions (Crenshaw et al. 1999, 
2003a). In this case, one would expect detectable (and likely saturated) 
columns of O~VI and H~I, which are not seen at the position of Component 1 (see 
Figure 3). The expected O~VI and Ly$\beta$ absorption lines are likely
``hidden'' by a substantial contribution from unabsorbed UV continuum emission, 
as demonstrated in \S5.3.

\subsection{Component 2}

It is instructive to compare our measurements of Component 2 with those of the 
Galactic component in Table 1. Since Component 2 is close to the systemic 
velocity (although with a slight blueshift), it could arise from the ISM of the 
host galaxy. However, an origin in the NLR or X-ray absorbers cannot be ruled 
out by the kinematics alone, since the radial velocity of Component 2 ($-$114 
km s$^{-1}$) is consistent with the small widths of the narrow emission lines, 
and we have no constraints on the kinematics of the X-ray absorbers.

We can obtain further insight by comparing the EWs of the lines from 
Component 2 with those of the Galaxy. In Table 1, the EWs of the low-ionization 
lines are very similar, whereas those of the higher ionization lines of Si~IV 
and C~IV are about half the values of their Galactic counterparts.
Component 2 could therefore arise from the ISM of NGC 4395, assuming it has a 
smaller amount of high-ionization gas than the Galactic column along 
this particular line of sight.
Moreover, we can rule out an origin in photoionized gas, regardless of whether 
it is the NLR, UV absorbers, or X-ray absorbers. The key is the ratio of
N(Si~IV)/N(C~IV) which is $\geq$ 1.5.
Given a power-law continuum and the abundances in \S5.3, we are unable to 
reproduce this ratio with photoionization models covering a broad range in 
ionization parameter and hydrogen column density.
Our models predict this ratio should always be $\leq$ 0.3, primarily due to the 
abundance 
ratio. However, N(Si~IV)/N(C~IV) ratios as high as $\sim$1 can be found in 
Galactic halo gas (Savage \& Sembach 1994), which is primarily collisionally 
ionized.

A final point in favor of an ISM origin is that the C~II and Ly$\beta$ lines 
for Component 2 in the FUSE spectrum are saturated near the zero flux level.
If these lines originated solely in either the NLR (with an angular size 
$<$0\arcsecpoint4, Filippenko et al. 1993) or absorbers near the nucleus, 
the extra far-UV flux in the large {\it FUSE} aperture would make them appear 
much shallower.

\subsection{Connection to the X-ray Absorbers}

We have generated photoionization models to further explore the possibility 
that the X-ray absorbers in NGC~4395 produce observable UV absorption lines.
A long (640 ks) {\it ASCA} observation from 
2000 revealed an underlying continuum characterized by a photon index $\Gamma$ 
$\approx$ 1.46, modified by a large column of gas intrinsic to NGC 4395, which 
SH2003 modeled as a two-zoned X-ray absorber. 
The zone closer to the nucleus was highly variable while the ionization state 
and total column density of the outer zone remained relatively constant during 
their observation.  Both of the components are highly ionized, and their 
observed effect is a depression of the continuum below $\sim$ 3 keV, due to the 
combined effects of the bound-free edges of species such as O~VII, O~VIII, 
Ne~IX, and Ne~X. Notably, in their 17 ks {\it Chandra} spectrum, Moran et al. 
(2004) also found evidence for complex absorption, although these data are of
insufficient quality to tightly constrain the physical characteristics of the
absorber.

The large column densities predicted by SH2003 suggest that 
detectable columns of H~I and O~VI may arise within the X-ray absorber. To 
calculate these column densities, we used the photoionization code 
CLOUDY90 (Ferland et al. 1998) to regenerate the constant component from
SH2003, which has a lower ionization than the variable component, and thus 
includes a larger fraction of the lower ionization species. For their models, 
SH2003 assumed that the X-ray continuum, with a spectral index $\alpha$ 
$=$ 0.46, extends to lower energies without turning up to meet the UV, which 
severely underpredicts the UV continuum. Therefore, we assumed a model 
spectral energy distribution (SED) closer to that which we derived for our 
emission-line study (Kraemer et al. 1999), with the form
F $\propto$ $\nu^{-\alpha}$, where $\alpha = 1.7$ for 13.6 $\leq {\rm h}\nu <$
1000 eV, and $\alpha = 0.46$ for ${\rm h}\nu \geq$ 1000 eV. 
SH2003 use the ionization parameter $\xi$ (in erg cm s$^{-1}$), which is 395.3 
$\times$ U for their adopted SED, where U is the number of ionizing photons 
per H atom. Based on our SED, we fixed the flux at 0.5 keV in order to predict 
the same H-like and He-like O and Ne columns, and hence the same opacity near 1 
keV. This results in a factor of 28.5 increase in U, due to the larger number 
of EUV photons. So for the constant component of the X-ray absorber modeled by 
SH2003, their value $\xi = 200$ corresponds to $U = 14.4$.

Assuming solar abundances, SH2003 derive a total hydrogen column density 
N$_{H} = 2.45 \times 10^{22}$ cm$^{-2}$, for the constant 
component of the absorber. However, in our emission-line study 
(Kraemer et al. 1999), we determined that the heavy element abundances in the 
NLR of NGC 4395 were sub-solar, with most elements $\sim$ 1/2 solar by number 
with respect to H, and
N/H $\approx$ 1/6 solar. Therefore, in order to reproduce the X-ray opacity 
from 
the SH2003 model, we
increased the total column density to N$_{H} = 4.0 \times 10^{22}$ cm$^{-2}$. 
We assumed that the X-ray absorber
is free of cosmic dust and all elements are in the gas phase.

In Table 2, we compare our predictions of ionic column densities from models 
using the SH2003 SED and our revised SED. The high-ionization columns derived 
from the X-ray observations are essentially the same. However, the H~I and O~VI 
column densities are factors of $\sim$17 and $\sim$4 smaller, respectively, in 
the model 
that uses our more realistic SED, due to the much larger flux of UV photons. 
Nevertheless, these columns are still quite large, and should be detectable in 
the {\it FUSE} band if the entire source of UV continuum emission is covered.
Our revised SED model predicts undetectable columns of C~IV (3.4 
$\times$ 10$^{11}$ cm$^{-2}$) and other low-ionization lines, so these lines in 
Components 1 and 2 are not from the X-ray absorbers.

So where are the expected H~I and O~VI absorption 
lines from the X-ray absorbers of SH2003? Given their column densities from our
revised SED model in Table 2, we have generated simulated absorption-line 
profiles to determine if they would be detectable in the {\it FUSE} spectra. 
For the purpose of illustration, we assumed that the X-ray component has the 
radial 
velocity and FWHM ($=$ 60 km s$^{-1}$) of Component 2. In Figure 3, we overplot 
the simulated absorption lines for two cases: complete covering of the UV 
continuum and 50\% covering in the line of sight. The latter comes from 
assuming that the broad-line emission is covered and that the excess continuum 
flux in the {\it FUSE} is not, resulting in a continuum covering factor of 0.33 
and an effective covering factor of $\sim$0.5 at the Component 2 positions. 
Figure 3 demonstrates that for complete covering, the 
Ly$\beta$ absorption from the X-ray absorber would be detectable, unless it is 
hidden in the low-ionization gas associated with Component 2 or filled in by 
geocoronal Ly$\beta$ emission. O~VI absorption at full covering would 
be detectable at Component 2 or any other velocity, unless it is fortuitously 
covered by the narrow O~VI emission at the systemic velocity of NGC 4395, in 
which case it would not be outflowing. For the case of 50\% covering, the H~I 
and O~VI absorption could be present at Component 2 or any other radial 
velocity, given the noise level of the {\it FUSE} spectra. Thus, although we do 
not detect the expected O~VI and H~I absorption lines, the {\it FUSE} spectra 
are not inconsistent with the parameters derived for the X-ray absorbers by 
SH2003.

\section{Conclusions}

We confirm the detection of C~IV absorption in NGC~4395, and have 
identified a number of other absorption lines in new high-resolution {\it FUSE} 
and {\it HST}/STIS spectra. In addition to Galactic lines, we have identified 
two components that are likely to be intrinsic to NGC~4395. Component 1 at 
$-$770 km s$^{-1}$ (with respect to $z =$ 0.0012) is only detected in the C~IV 
$\lambda$1548.2 line. If it is real, it cannot arise in the ISM or NLR of 
NGC~4395 due to its high velocity. Furthermore, the X-ray absorbers 
characterized by SH2003 are too highly ionized to produce a detectable C~IV 
column. Thus, Component 1 likely originates in an outflowing UV absorber 
similar to those seen in most Seyfert 1 galaxies (Crenshaw et al. 2003a).
Observations at higher S/N are required to confirm the reality of this 
component, and identify additional lines (e.g., N~V $\lambda\lambda$1238.8, 
1242.8) to characterize its physical conditions.

The possibility that Component 1 arises from UV absorbers close to the nucleus, 
for example in an accretion-disk wind (e.g., Proga 2003), remains an intriguing 
prospect, given the low luminosity of the central continuum source. Drawing 
on the analogy with UV absorbers in normal Seyfert 1s, possible indicators of a 
location close to the nucleus would be variability in the absorption lines or 
partial covering of the continuum and/or BLR. These measurements require 
high-resolution spectra at higher S/N than the current data.

Component 2 shows a broad range in ionization, and likely arises in the ISM of 
NGC~4395, since photoionization models cannot match the observed 
N(Si~IV)/N(C~IV) ratio, and the C~II and Ly$\beta$ absorption lines absorb all 
of the far-UV flux in the large (30$''$ $\times$ 30$''$) {\it FUSE} aperture. 
It is interesting that this component shows a slight outflow ($-$114 km 
s$^{-1}$) with respect to the host galaxy. 
This may indicate a large-scale galactic outflow driven by stellar winds and/or 
supernovae (Veilleux et al. 2002) near the center of the galaxy.

We have not detected the NLR in absorption, despite its large derived covering 
factor (Kraemer et al. 1999). This could be due to the overall geometry of the 
NLR or to clumpiness on a smaller scale (i.e., there is no NLR ``cloud'' in the 
line of sight to the continuum source).  However, we note that absorption from 
the ``OUTER'' component of the NLR in Kraemer et al. could be hidden in 
Component 2 (the other NLR components have saturated C~IV columns) if it shares 
the same outflow velocity as the ISM. We also do not detect the expected 
columns of O~VI and Ly$\beta$ absorption from the X-ray absorbers claimed by 
SH2003. 
This is likely due to a large contribution of hot stars to the far-UV flux in 
the {\it FUSE} aperture, although X-ray observations at higher S/N and spectral 
resolution would be helpful for confirming the presence and properties of the 
X-ray absorption.

\acknowledgments

We thank D. Shih, K. Iwasawa, and E. Moran for helpful discussions.
SBK and DMC acknowledge support from NASA grants HST GO-09362.01 and NAG5-13049.
Support for proposal GO-9362 was provided by NASA through a grant from the 
Space Telescope Science Institute, which is operated by the Association of 
Universities for Research in Astronomy, Inc., under NASA contract NAS 5-26555. 
The National Radio Astronomy Observatory is a facility of the National
Science Foundation, operated under cooperative agreement by Associated
Universities, Inc.

\clearpage

\figcaption[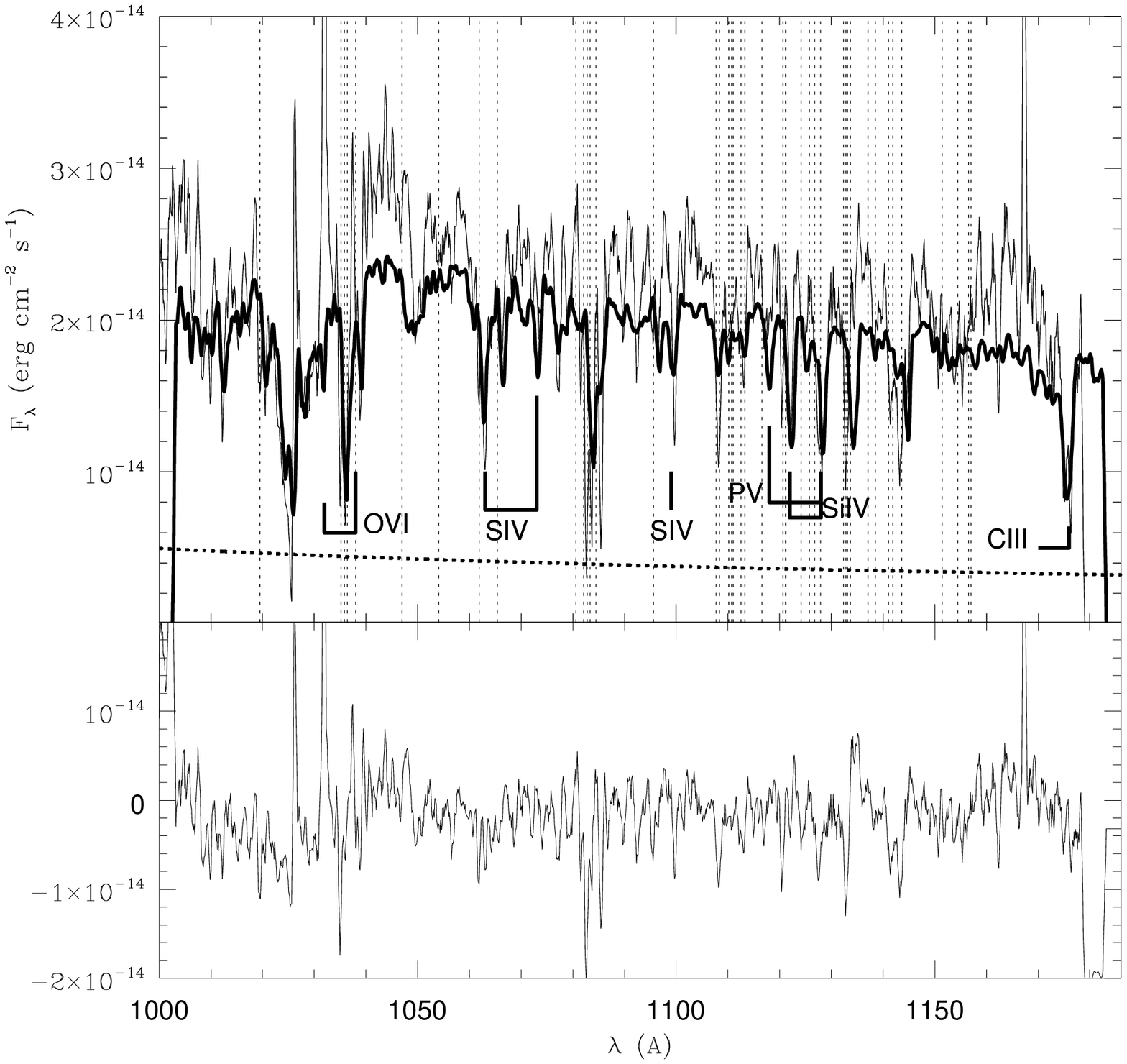]{Reddening and redshift corrected {\it FUSE} spectrum of NGC 
4395 (top panel) smoothed with a 5 point boxcar (thin line), the extention of 
the nuclear 
continuum to shorter wavelengths determined from the HST spectrum (thick dotted 
line), and the starburst model for an instantaneous burst 5~Myr old with 
metallicity 0.4 Z$_{\odot}$ and Salpeter IMF (thick solid line). The starburst 
spectrum is scaled so that, when added to the power law spectrum, the flux 
corresponds to that of NGC 4395 in the 1102 -- 1107 \AA\ range.
This figure shows only the wavelength region covered by the models.
The vertical dotted lines indicate interstellar absorption lines, from
our Galaxy, obtained from the list of Pellerin et al. (2002). The most 
prominent stellar lines from the 
list of Pellerin et al. (2002) are indicated under the spectrum with thick bars.
A number of H$_2$ absorption lines, not shown here,
also contribute to absorption lines seen in this spectrum.
The bottom panel of the figure shows the residuals from
the subtraction of the starburst and power law spectrum from the spectrum
of NGC 4395.}

\figcaption[f1.eps]{Portions of the STIS echelle spectra of NGC~4395, showing 
the absorption lines in different ions. The fluxes have been smoothed with a 
seven-point boxcar, and are plotted as a function of the radial velocity (of 
the strongest member for each doublet), relative to an emission-line redshift 
of $z =$ 0.0012. The kinematic components of the lines (including doublets) are 
numbered, and strong Galactic absorption lines are labeled with ``G''. The 
narrow components of the C~IV emission are labeled with ``n''. The 1992 {\it 
HST}/FOS spectrum is overplotted in the top panel in red.}

\figcaption[f1.eps]{Portions of the {\it FUSE} spectra of NGC~4395, plotted as 
in Figure 1. The absorption features in the region of O~VI $\lambda$1037.6 are 
actually due to C~II $\lambda$1036.3, as shown in the bottom panel. The narrow 
components of O~VI emission and geocoronal Ly$\beta$ emission are labeled with 
``n''. The simulated profiles described in \S5.3 are plotted for 50\% (blue) 
and 100\% (red) covering of the continuum plus line emission.}

\clearpage
\begin{deluxetable}{lccccc}
\tablecolumns{6}
\footnotesize
\tablecaption{Measurements of Absorption Components in NGC~4395}
\tablewidth{0pt}
\tablehead{
\colhead{Line} & \colhead{EW(G)$^a$} & \colhead{$v_r$(G)$^b$} &
\colhead{EW(2)$^a$} & \colhead{$v_r$(2)$^b$} & \colhead{N$_{ion}$(2)$^c$} \\
\colhead{} & \colhead{(\AA)} & \colhead{(km s$^{-1}$)} &
\colhead{(\AA)} &\colhead{(km s$^{-1}$)} & \colhead{(10$^{14}$ cm$^{-2}$)}
}
\startdata
Ly$\beta$ $\lambda$1025.7 & ---         & --- & 0.86 ($\pm$0.12) &$-$85   &
$>$ 11.7 \\
C~II $\lambda$1036.3  &0.48 ($\pm$0.06) &$-$37 &0.61 ($\pm$0.10) &$-$104  &
$>$5.13\\
O~I $\lambda$1302.2  &0.52 ($\pm$0.11) &$-$34 &0.42 ($\pm$0.12) &$-$124  &
$>$5.76 \\
C~II $\lambda$1334.5  &0.91 ($\pm$0.12) &$-$43 &0.89 ($\pm$0.10) &$-$115  &
---\\
Si~IV $\lambda$1393.8 &1.18 ($\pm$0.19) &$-$18 &0.43 ($\pm$0.11) &$-$96   &
---\\
Si~IV $\lambda$1402.8 &---              &---   &0.55 ($\pm$0.18) &$-$94   &
$>$ 1.21 \\
C~IV $\lambda$1548.2  &0.52 ($\pm$0.07) &$-$16 &0.25 ($\pm$0.09) &$-$135  &
0.82 ($\pm$0.29) \\
C~IV $\lambda$1550.8  &---              &---   &0.21 ($\pm$0.09) &$-$155  &
---\\
Mg~II $\lambda$2796.3 &1.32 ($\pm$0.27) &$-$33 &1.63 ($\pm$0.19) &$-$113  &
---\\
Mg~II $\lambda$2803.5 &0.92 ($\pm$0.19) &$-$23 &1.49 ($\pm$0.24) &$-$118 &
$>$0.73
\enddata
\tablenotetext{a}{Equivalent widths for kinematic Components ``G'' (Galactic) 
and ``2''. See \S4 for Component 1 C~IV values.}
\tablenotetext{b}{Radial velocity centroid relative to $z =$ 0 for Component G 
and $z =$ 0.0012 for Component 2. The uncertainty in these measurements is 
about 
$\pm$ 10 km s$^{-1}$.}
\tablenotetext{c}{Column density from the line that gives the 
tightest constraint for each ion.}
\end{deluxetable}

\clearpage
\begin{deluxetable}{lcc}
\tablecolumns{3}
\footnotesize
\tablecaption{Warm Absorber Columns$^a$}
\tablewidth{0pt}
\tablehead{
\colhead{Ion} & \colhead{SH2003 SED} & \colhead{Revised SED}\\
\colhead{} & {(cm$^{-2}$)} & {(cm$^{-2}$)}
}
\startdata
H~I & 2.6 $\times$ 10$^{16}$ & 1.5 $\times$ 10$^{15}$ \\
O~VI & 3.9 $\times$ 10$^{15}$ & 1.1 $\times$ 10$^{15}$ \\
O~VII & 5.0 $\times$ 10$^{17}$ & 5.3 $\times$ 10$^{17}$ \\
O~VIII & 4.9 $\times$ 10$^{18}$ & 4.6 $\times$ 10$^{18}$ \\
Ne~IX & 5.1 $\times$ 10$^{17}$ & 5.7 $\times$ 10$^{17}$ \\
Ne~X & 1.3 $\times$ 10$^{18}$ & 1.2 $\times$ 10$^{18}$ \\
\enddata
\tablenotetext{a}{Predicted ionic column densities for the constant warm 
absorber of SH2003.}
\end{deluxetable}

\clearpage
\begin{figure}
\epsscale{0.9}
\plotone{f1.eps}
\\Fig.~1.
\end{figure}

\clearpage
\begin{figure}
\epsscale{0.9}
\plotone{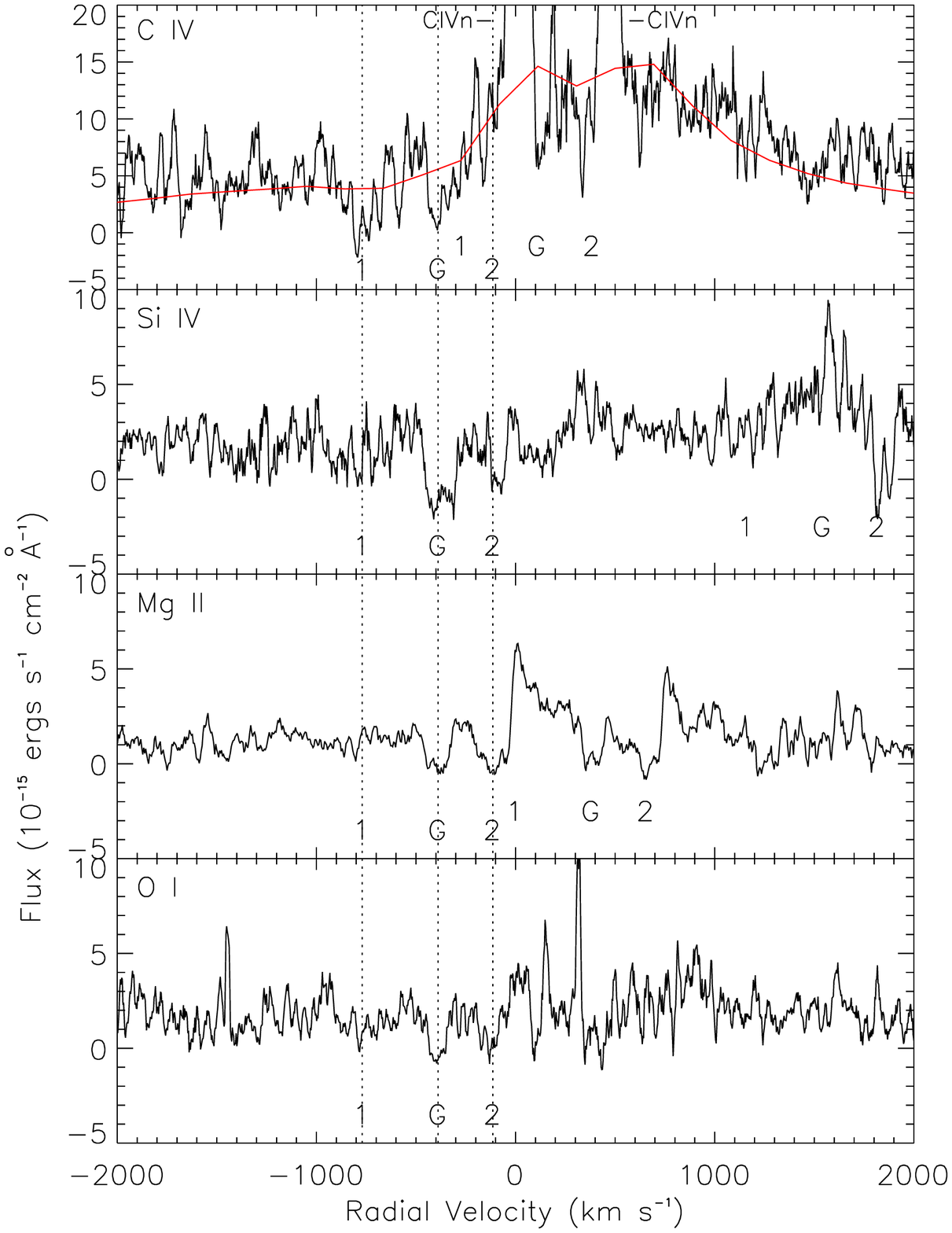}
\\Fig.~2.
\end{figure}

\clearpage
\begin{figure}
\epsscale{0.9}
\plotone{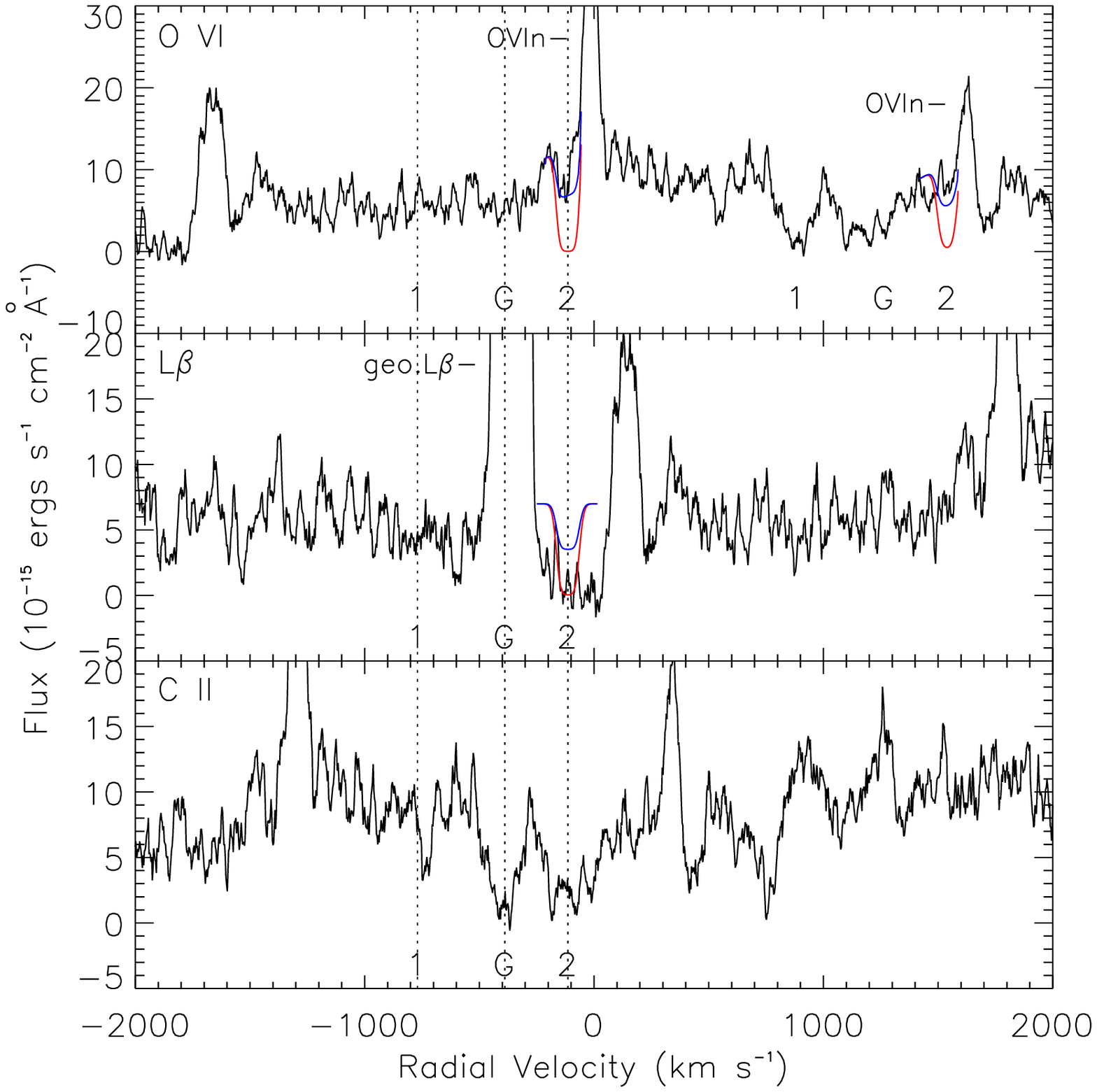}
\\Fig.~3.
\end{figure}

\end{document}